\documentstyle[twoside,fleqn,npb,epsfig]{article}
\newcommand{\bea}{\begin{eqnarray}}
\newcommand{\bq}{\begin{equation}}
\newcommand{\eea}{\end{eqnarray}}
\newcommand{\eq}{\end{equation}}

\newcommand\xx{\tilde{x}}

\title{Representation of the virtual Compton amplitude for polarized 
scattering in the generalized Bjorken region\thanks{Contribution to
the Proceedings of the 7th International Workshop on Deep Inelastic 
Scattering
and QCD, Zeuthen, April 1999, Nucl. Phys. {\bf B} (Proc. Suppl.)}}

\author{Johannes Bl\"umlein\thanks{Supported in part by EU contact
FMRX--CT98--0194}
\address {DESY Zeuthen, Platanenallee 6, D--15735 Zeuthen, Germany},
Bodo Geyer\address{Naturwiss.--Theoretisches Zentrum,
Universit\"at Leipzig, Augustusplatz 10, D--04109 Leipzig, Germany}
and
Dieter Robaschik$^{a}$}

\begin{document}

\vspace{-3mm}
\begin{abstract}
The Compton amplitude is calculated in terms of expectation values of 
light--ray quark operators. As a technical tool we apply the nonlocal
light--cone expansion. Thereby we express the expectation value of the
vector light--ray operator with the help of the expectation value of the
corresponding scalar operator of twist 2. This allows important 
simplifications. In the limit of forward scattering the integral 
relations between the twist--2 contributions of the structure functions
are implied directly.
\end{abstract}

\maketitle

\vspace{-3mm}
The Compton amplitude for the scattering of a virtual photon off a hadron
$\gamma^* + p_1 \rightarrow \gamma^{'*} + p_2 $ provides one of the basic
tools to understand the short--distance behavior of the nucleon and to
test Quantum Chromodynamics (QCD) at large space--like virtualities
\cite{ZEUT}. The Compton amplitude for the general case of nonforward
scattering is given by
\begin{eqnarray}
\label{COMP}
\lefteqn
{T_{\mu\nu}(p_+,p_-,q)
= i \int d^4x \,e^{iqx} \,
}
 \\
& & \langle p_2, S_2\,|T (J_{\mu}(x/2) J_{\nu}(-x/2))|\,p_1,
S_1\rangle~.\nonumber
\end{eqnarray}
Here, $ p_+ = p_2 + p_1,~q= (q_1 + q_2)/2,  p_- = p_2 - p_1$,       where
$q_1 (q_2)$ and $p_1 (p_2)$ denote the four--momenta of the incoming
(outgoing) photon and hadron, respectively, and $S_{1}, S_{2}$ are the 
spins of the initial-- and final--state hadron. The {\it generalized
Bjorken region} is defined by the conditions
$\nu =  qp_+ \rightarrow \infty, Q^2 = - q^2 \rightarrow \infty~,$ 
keeping the variables $\xi  =  {Q^2 }/{qp_+}$ and $\eta = {qp_-}/{qp_+} 
= (q_1^2 - q_2^2)/{2\nu}$ fixed. The (renormalized) time--ordered product
in Eq.~(\ref{COMP}) can be represented in terms of the operator product 
expansion. The non--local operator product expansion~[1--3] leads to  
compact expressions for the coefficient functions and the operators in 
the nonforward case. In lowest order in the coupling constant the
expansion contains only quark operators with two external legs. The 
electromagnetic current reads
$J_\mu(x) = \overline \psi(x) \gamma_\mu \lambda^{\rm em} \psi(x),$
where $\psi(x)$ are the quark fields and $\lambda^{\rm em}$ projects onto
the flavor states. One obtains:
\begin{eqnarray}
\label{tpro}
\lefteqn{
T(J_{\mu}(x/2) J_{\nu}(-x/2) ) \approx
\int_{-1}^{+1} d \kappa_+\int_{-1}^{+1} d \kappa_- }
\\
\hspace{-1.5cm}
& & \times \Bigl[C_a(x^2,\kappa_{\pm},\mu^2)
\Bigl(g_{\mu \nu} O^a (\kappa_{\pm} \xx \mu^2)
\nonumber \\
& & -\xx_{\mu} O^a_\nu (\kappa_{\pm} \xx, \mu^2)
 -\xx_{\nu} O^a_\mu (\kappa_{\pm} \xx, \mu^2)
\Bigl)
\nonumber\\
& &
+ \,i\,C_{a,5}(x^2,\kappa_{\pm},\mu^2)
{\varepsilon_{\mu\nu}}^{\rho\sigma}
\xx_{\rho} O_{5, \sigma}^a (\kappa_{\pm} \xx, \mu^2)
 \Bigr].\nonumber
\end{eqnarray}
For convenience we introduced auxiliary integrations over the variables 
$\kappa_\pm $, and a light--ray vector  $\tilde x$  corresponding to
the vector $x$. Here $C_{a (5)}$ are the 
renormalized coefficient functions which are used in Born approximation,
and $O^a_{(5)\mu}(\kappa_1 \tilde x, \kappa_2 \tilde x)$
are the renormalized (anti)symmetric light--cone operators with
$\kappa_{\pm} = (\kappa_2 \pm \kappa_1)/2$,
\begin{eqnarray}
\label{qop}
 O^a_\mu
&=&
\frac{1}{2}[
\overline\psi(\kappa_1\xx)\lambda^a_f\gamma_\mu
 \psi(\kappa_2\xx)- (\kappa_1 \leftrightarrow \kappa_2)
], \nonumber \\
 O^a_{5,\mu}
&=&
\frac{1}{2}
[
\overline\psi(\kappa_1 \xx)\lambda^a_f\gamma_5\gamma_\mu
 \psi(\kappa_2\xx)
+(\kappa_1 \leftrightarrow \kappa_2)], \nonumber
\end{eqnarray}
respectively. The phase factors drop out in the light--cone gauge 
$\tilde x A =0.$
We consider first the quark operators $O^a_\mu$ and
$O_{5, \mu}^{a}$, which contain contributions of twist--2, 3 and 4.
The explicit computations  lead to the following expressions for the 
twist--2 light--ray vector operators \cite{GLR} in terms of the 
corresponding scalar operator
\begin{eqnarray}
\label{optw2}
O_{\mu}^{q,\, \rm tw2} (- \kappa \xx, \kappa \xx)
 = \! \int_0^{1}\!\! d {\tau}
\partial_\mu \!
\left.
O^q_{\rm trl}(-\kappa\tau x, \kappa\tau x)
\right|_{x \rightarrow \xx}.\nonumber
\end{eqnarray}
Let us first consider the matrix elements of the the scalar twist--2 
quark operator
\begin{eqnarray}
\label{kdec1}
\lefteqn{
\langle p_2|O^{q}_{\rm trl }(-\kappa_- x, \kappa_- x)|p_1\rangle
= } \\
& &\tilde g^q(\kappa_- xp_\pm, \kappa_-^2 x^2)
\overline u( p_2) (\gamma x) u( p_1) \nonumber  \\
&+& \tilde h^q(\kappa_- xp_\pm, \kappa_-^2 x^2)
\overline u( p_2)x (\sigma p_-)  u( p_1)/M. \nonumber
\end{eqnarray}
Using the foregoing relation  we get an expression for the matrix element
of the vector operator,
\begin{eqnarray}
\label{kdecv}
\lefteqn{
\langle p_2|O^{q,\;\rm tw2}_\mu
(\kappa_1 \xx, \kappa_2 \xx)|p_1\rangle } \\
\hspace{-0.3cm}
&= &
\int_0^1 \! d \lambda \, \partial_\mu^x \,
\langle p_2|O^{q}_{\rm trl}(\kappa_1 \lambda x,
\kappa_2 \lambda x)|p_1\rangle\Big|_{x=\xx}
\nonumber
\\
&=&
\int_0^1 \! d \lambda \,
\partial_\mu^x \Bigl\{
 e^{i\kappa_+\lambda xp_-} \Bigl[
\overline u(p_2) (\gamma x) u(p_1)\nonumber \\
& &\times
 \tilde g^q(\kappa_- \lambda x p_+,\kappa_- \lambda x p_-,
\kappa^2_- \lambda^2 x^2) 
\nonumber \\
& &+
\overline u(p_2)x (\sigma p_-)  u(p_1)/M \nonumber \\
& & \times
\tilde h^q(\kappa_- \lambda x p_+,\kappa_- \lambda x p_-,
\kappa^2_- \lambda^2 x^2)\Bigr]\Bigr\}\nonumber
\Big|_{x=\xx}.
\nonumber
\end{eqnarray}
We now apply the Fourier transformation  to $\tilde g (\tilde h)$
with measure $Dz    =    dz_+ dz_-~~\theta(1+ z_+ + z_-) \newline
\theta(1+ z_+ - z_-) \theta(1- z_+ + z_-) \theta(1- z_+ - z_-):$
\begin{eqnarray}
\lefteqn{
\label{zrep}
\tilde f(\kappa x p_+, \kappa x p_-, \kappa^2 x^2)
 = }  \\
& &
\int Dz
e^{-i\kappa x(p_+z_+ + p_- z_-)} f(z_+,z_-, \kappa^2 x^2)~,
\nonumber
\end{eqnarray}
and perform  the $\lambda$--integration in Eq.~(\ref{kdecv}), which 
yields
\begin{eqnarray}
\label{MAIN}
\lefteqn{
\langle p_2|O^{q,\;\rm twist 2}_\mu
(\kappa_1 \xx, \kappa_2\xx)|p_1\rangle } \nonumber \\
& = &\int Dz
e^{-i\kappa_- \xx (p_+ z_+ + p_- z_- }
\Bigl\{
G^q(z_+,z_-)\nonumber \\
& & (\overline u(p_2)\gamma_{\mu}u(p_1)
 - i\kappa_- p_\mu(z)
\overline u(p_2) (\gamma x) u(p_1))
\nonumber\\
& & +
\overline u\gamma_{\mu}u
[(i\kappa_+ p_-^{\mu} \partial_{i\kappa_+ x p_-}
+ 2 x_\mu \kappa^2_-\partial_{\kappa^2_- x^2})]
\nonumber \\
& & \times G^q(z_+,z_-,\kappa_+ x p_-, \kappa^2_- x^2)
\Bigr\}\Big|_{x=\xx}
\nonumber \\
& & + \hbox{similar terms containing $ H^q $ }.
\nonumber
\end{eqnarray}
The functions $G^q (H^q)$ read
\begin{eqnarray}
\lefteqn{
G(z_+,z_-, \kappa_+ xp_-,\kappa^2_- x^2)
=\int_0^1 \frac{d\lambda}{\lambda^2}
e^{i \kappa_+ \lambda x p_-}
}\nonumber \\
& & \times
g\Bigl(\frac{z_+}{\lambda},\frac{z_-}{\lambda},
\kappa^2_- \lambda^2 x^2\Bigr)
\theta (\lambda - |z_+|)
\theta (\lambda - |z_-|), \nonumber
\end{eqnarray}
\begin{eqnarray}
\label{tt11}
\stackrel{o}{G}
&=&
\partial_{i\kappa_+ x p_-}
G(z_+,z_-, \kappa_+ xp_-,\kappa^2_- x^2)\nonumber\\
&=&
\int_0^1 \frac{d\lambda}{\lambda}
 g\Bigl(\frac{z_+}{\lambda},\frac{z_-}{\lambda} ,
\kappa^2_- \lambda^2 x^2\Bigr)
 e^{i \kappa_+ \lambda x p_-} \nonumber \\
& &  \times \theta (\lambda - |z_+|) \theta (\lambda - |z_-|),
\nonumber\\
G'
&=&
\partial_{\kappa^2_- x^2 }
G(z_+,z_-,\kappa_+ xp_-, \kappa^2_- x^2) \nonumber \\
&=&\int_0^1 d\lambda\,
\partial_{\kappa^2_- \lambda^2 x^2 }
g\Bigl({z_+}{\lambda},\frac{z_-}{\lambda}
\kappa^2_- \lambda^2 x^2\Bigr)\nonumber \\
& & \times
\theta (\lambda - |z_+|)
\theta (\lambda - |z_-|)  e^{i \kappa_+ \lambda x p_-}~.
\nonumber
\end{eqnarray}
Consequences of this representation are the integral relations in
polarized deeply inelastic scattering in the forward case~\cite{BLK}.
Analogous decompositions and representations are valid for the matrix
elements of the pseudo--scalar and pseudo--vector twist--2 quark 
operators. The corresponding partition functions are denoted by
$(g^q_5, \, h^q_5)$ and  $(G^q_5, \, H^q_5)$, respectively, and the 
Dirac and Pauli structures are to be replaced substituting 
$\gamma_{\mu} \rightarrow \gamma_5 \gamma_{\mu}$ and
$\sigma_{\mu\nu} \rightarrow \gamma_5 \sigma_{\mu\nu}$.
With the prerequisites provided above we now derive the asymptotic 
representation of the Compton scattering amplitude in the generalized 
Bjorken region, noting that $\xi = Q^2/qp_+, \eta = {qp_-}/{qp_+}$,
\begin{eqnarray}
\label{COMPAS}
\lefteqn{
T_{\mu \nu}^{\rm tw2}(p_+,p_-,q)
=
\int Dz \bigg\{  } \\
& &
\bigg(\frac{1}{\xi + t -i\varepsilon} -
      \frac{1}{\xi - t  -i\varepsilon}\bigg)
 F^{(1)}_{\mu \nu}
 \nonumber\\
& &  +\;
\bigg(\frac{1}{(\xi + t  -i\varepsilon)^2} +
      \frac{1}{(\xi - t  -i\varepsilon)^2}\bigg)
 F^{(2)}_{\mu \nu}
\nonumber \\
& &  +\;
\bigg(\frac{1}{(\xi + t -i \varepsilon)^3} -
      \frac{1}{(\xi - t -i\varepsilon)^3}\bigg)
F^{(3)}_{\mu \nu}
\nonumber \\
& &  +\;
\bigg(\frac{1}{\xi + t -i \varepsilon} +
      \frac{1}{\xi - t -i\varepsilon}\bigg)
F^{(1)}_{5,\mu \nu}
\nonumber \\
& &  +\;
\bigg(\frac{1}{(\xi + t -i\varepsilon)^2} -
      \frac{1}{(\xi - t -i\varepsilon)^2}\bigg)
F^{(2)}_{5,\mu \nu}
\bigg\}.
\nonumber
\end{eqnarray}
With $ t= z_+ + \eta z_- $ Eq.~(\ref{COMPAS}) is the two--variable
representation of the nonforward Compton amplitude. One may generalize 
Eq.~(\ref{COMPAS}) accounting for current conservation explicitly. The 
functions  $F^{(k)}_{(5)\mu\nu}(p_+, p_-, q; z_+, z_-)$ are
(anti)symmetric w.r.t.~the Lorentz indices and contain the functions 
$G, H$, momentum vectors and the spinor structures $\bar u \gamma^\mu u$ 
etc. As an example we present one of these functions here \cite{ZEUT}:
\begin{eqnarray}
F_{\mu \nu}^{(1)}
&= &
\frac{1}{qp_+}
\bigg[q^\alpha g_{\mu\nu}
- (g^\alpha_{\;\nu} q_\mu + g^\alpha_{\;\mu} q_\nu)\bigg] \\
&\times & [ \bar u(p_2)\gamma_\alpha  u(p_1) G^q
\nonumber\\ & &
+ \Bigl(
\bar u(p_2)(\sigma p_-)^\alpha u(p_1)/M \Bigr)
H^q ]. \nonumber
\end{eqnarray}
It is, however, also possible to derive the one--variable representation
from Eq.~(6). In this case we use instead of the variables $z_+, z_-$ the
variables $ t , z_- $. Then the integrations factorize. The
$t$--integration remains as overall 
integration and the $z_-$--integration
has to be
performed in the  functions $F^{(i)}_{(5),\mu \nu}$ which depend 
on the functions $G, H$. Finally the $z_-$--integration acts on the
functions $G^q_{(5)} (H^q_{(5)})$ only, leading to new functions
\begin{eqnarray}
\label{x3}
\lefteqn{
\widehat F^q_{(5)}(t,\eta) =
\int_{-1}^{+1} d z_- 
F^q_{(5)}(z_+ =t-\eta z_- ,z_-)
}\nonumber \\
& &
\times
\theta(1 + t - \eta z_- + z_-)
\theta(1 + t - \eta z_- - z_-) \nonumber\\ & & \times
\theta(1 - t + \eta z_- + z_-)
\theta(1 - t + \eta z_- - z_-), \nonumber
\end{eqnarray}
with $F = G$ or $H$,
which depend on the variables $t$ and $\eta$. In the limit of forward 
scattering, $p_- \rightarrow 0$, the Compton amplitude does not depend on
the distribution amplitudes $H^q_{(5)}(z_+,z_-)$ any longer.  Using the
normalizations
${\bar u}(p) u(p)  = 2M , {\bar u}(p)\gamma_\mu u(p)  = 2p_\mu$ and
$ {\bar u}(p)\gamma_5\gamma_\beta u(p)  = -2 S_\beta, S^2 = - M^2~$,
we obtain
\begin{eqnarray}
\label{CV3}
\lefteqn{
T_{\mu \nu}^{\rm sym} =
\left(g^{\mu\nu} - \frac{p^\mu q^\nu + p^\nu q^\mu}{pq}\right)
 \int_{-1}^1 dz_+ } \\
& &\times
\left(\frac{1}{\xi + z_+ -i\varepsilon}
     -\frac{1}{\xi - z_+ -i\varepsilon}\right)
      \widehat{g}^q(z_+) \nonumber\\
&+& 2 q_{\mu} q_{\nu} \int Dz
 {G'}^q(z_+,z_-)\nonumber \\
& & \times
\left(
\frac{1}{(\xi + z_+ - i \varepsilon)^3}
-\frac{1}{(\xi - z_+ - i \varepsilon)^3}\right),\nonumber
\end{eqnarray}
\begin{eqnarray}
\label{x4}
\lefteqn{
T_{\mu \nu}^{\rm antisym}
= i
   {\varepsilon_{\mu\nu}}^{\gamma\beta}
   \frac{q_{\gamma}p_{\beta}}{(pq)^2}qS
 \int_{-1}^{+1} dz_+ } \\
& & \times
\left[\frac{1}{\xi + z_+ - i\varepsilon}
+ \frac{1}{\xi - z_+ -i \varepsilon}\right]\nonumber \\
& &  \times
\left[ \int_{z_+}^{{\epsilon}(z_+)} \frac{dz}{z}
\widehat{g}^q_5(z) - \widehat{g}^q_5(z_+)\right]
\nonumber\\ & &
-i {\varepsilon_{\mu\nu}}^{\gamma\beta} \frac{q_{\gamma}S_{\beta}}{(pq)  }
   \int_{-1}^{+1} dz_+ \nonumber \\
& & \hspace*{-6mm}
\times
\left[\frac{1}{\xi + z_+ - i\varepsilon} + \frac{1}{\xi - z_+
-i \varepsilon}\right]
 \int_{z_+}^{ \epsilon(z_+)} \frac{dz}{z}
\widehat{g}^q_5(z).\nonumber
\end{eqnarray}
Here, the $z_-$--integral was performed
\begin{eqnarray}
\label{CV2}
\int^{+1-|z_+|}_{-1+|z_+|} dz_- G^q(z_+,z_-)
=  \int_{ z_+ }^{{\rm \epsilon}(z_+)}       \frac{dz'}{z'}
\widehat{g}^q(z'), \nonumber
\end{eqnarray}
with $\epsilon(z) = {\rm sign}(z)$ and
\begin{eqnarray}
\label{}
\widehat g^q(z_+) = \int_{-1 - |z_+|}^{1- |z_+|} dz_- g^q(z_+,z_-).
\nonumber
\end{eqnarray}
From the absorptive part of the Compton amplitude in the forward
direction one derives directly
the Wandzura--Wilczek relation \cite{BLK}. 
Moreover the tensor structure of Eq.~(\ref{CV3})
implies the Callan--Gross relation. Similarly the other twist--2 relations
are obtained~\cite{BLK}.
Note that the partition functions
$G,H$ and $G_5, H_5$ are directly related to the partition functions 
$g, h$ and $g_5, h_5$ which are defined by the matrix elements of scalar 
and pseudo--scalar twist--2 quark operators, Eq.~(\ref{kdec1}). This 
demonstrates again that for the description of virtual Compton
scattering the properties of the scalar operators are sufficient in 
leading order.


\begin{thebibliography}{9}
%
\bibitem{ZEUT}
J. Bl\"umlein, B. Geyer, D. Robaschik, Phys. Lett. {\bf B 406} (1997) 161
and Erratum; DESY 99--020, {\tt hep-ph/9903520}.
%
\bibitem{AS}
S.A. Anikin and O.I. Zavialov, Ann. Phys. (NY) {\bf 116} (1978) 135.
%
\bibitem{SLAC}
B.~Geyer, D.~M\"uller, and D.~Robaschik, preprint SLAC-Pub-6495 (1994).
%
\bibitem{GLR}
B. Geyer, M.~Lazar, and D. Robaschik, {\tt hep-th/9901090}.
\bibitem{BLK}
C.G. Callan and D.J. Gross, Phys. Rev. Lett. {\bf 22} (1969) 156;
D. Dicus, Phys. Rev. {\bf D5} (1972) 1367;
S. Wandzura and F.Wilczek, Phys. Lett. {\bf B 72} (1977) 95;
J. Bl\"umlein and N. Kochelev, Phys. Lett. {\bf B381} (1996) 296;
Nucl. Phys. {\bf B498} (1997) 285; J.~Bl\"umlein and A. Tkabladze,
DESY 98/181,
{\tt hep-ph/9812478}, Nucl. Phys. {\bf B} in print.
\end{thebibliography}
\end{document}